\documentclass[unicode,
reprint,
superscriptaddress,
amsmath,amssymb,
aps,
prb,
]{revtex4-2}

\usepackage{graphicx}
\usepackage{dcolumn}
\usepackage{bm}
\usepackage{times}
\usepackage{orcidlink}

\hypersetup{
   plainpages=false,
   colorlinks=true,       
   citecolor=blue,        
}
\usepackage{xcolor}

\begin{document}

\title{Bound excited states of Fr\"ohlich polarons in one dimension}

\author{J. Taylor \orcidlink{0009-0009-3756-0566}}
\affiliation{Dodd-Walls Centre for Photonic and Quantum
Technologies, Auckland 0745, New Zealand}
\affiliation{Centre for Theoretical Chemistry and
Physics, New Zealand Institute for Advanced Study,
Massey University, Private Bag 102904, North Shore,
Auckland 0745, New Zealand}
\affiliation{%
MacDiarmid Institute for Advanced Materials and Nanotechnology, Auckland 1010, New Zealand
}%
\affiliation{Department of Physics, University of Auckland, Auckland 1010, New Zealand}

\author{M. Čufar \orcidlink{0000-0003-0734-2719}}%
\affiliation{Dodd-Walls Centre for Photonic and Quantum
Technologies, Auckland 0745, New Zealand}
\affiliation{Centre for Theoretical Chemistry and
Physics, New Zealand Institute for Advanced Study,
Massey University, Private Bag 102904, North Shore,
Auckland 0745, New Zealand}

\author{D. Mitrouskas \orcidlink{0000-0002-8058-3263}}
\author{R. Seiringer \orcidlink{0000-0002-6781-0521}}
\affiliation{
Institute of Science and Technology Austria (ISTA), Am Campus 1, 3400 Klosterneuburg, Austria
}%
\author{E. Pahl \orcidlink{0000-0002-6685-1655}}
\affiliation{%
MacDiarmid Institute for Advanced Materials and Nanotechnology, Auckland 1010, New Zealand
}%
\affiliation{Department of Physics, University of Auckland, Auckland 1010, New Zealand}
\author{J. Brand \orcidlink{0000-0001-7773-6292}}
\email{j.brand@massey.ac.nz}
\affiliation{Dodd-Walls Centre for Photonic and Quantum
Technologies, Auckland 0745, New Zealand}
\affiliation{Centre for Theoretical Chemistry and
Physics, New Zealand Institute for Advanced Study,
Massey University, Private Bag 102904, North Shore,
Auckland 0745, New Zealand}

\date{\today}

\begin{abstract}
The one-dimensional Fröhlich model describing the motion of a single electron interacting with optical phonons is a paradigmatic model of quantum many-body physics. 
We predict the existence of an arbitrarily large number of bound excited states in the strong coupling limit and calculate their excitation energies. 
Numerical simulations of a discretized model demonstrate the complete amelioration of the projector Monte Carlo sign problem by walker annihilation in an infinite Hilbert space. They 
reveal the threshold for the occurrence of the first bound excited states at a value of $\alpha \approx 1.73$ for the dimensionless coupling constant. This puts the threshold into the regime of intermediate interaction strength. We find a significant spectral weight and increased phonon number of the bound excited state at threshold.
\end{abstract}

\maketitle

\section{Introduction}
Studies of the polaron, describing the motion of an electron in a polarizable medium, go back almost a century to the works of Landau \cite{Landau1933} and Pekar \cite{Pekar1954}. Formalised by Fröhlich \cite{Fröhlich1954a}, the polaron model is one of the most studied
problems of quantum many-body theory \cite{Devreese2009a} with real-world relevance for electronic material properties \cite{Franchini2021}. 
Calculations of the ground state energy, the dispersion relation and also the structure of the excitation continuum have been achieved at great accuracy and become an important test bed for analytical \cite{Feynman1955} and numerical \cite{Prokof'ev1998a,Mischenko2000a,Titantah2001,Hahn2018a} techniques. 

%
%
%
The existence of bound excited states below the excitation continuum in the strong coupling limit of the three-dimensional model was first predicted in the 1960s \cite{Mel'nikov1969,Levinson1974}, and rigorously proven very recently \cite{Mitrouskas2023a}.
Previous numerical studies have failed to produce evidence for bound excited states in the three-dimensional case \cite{Mischenko2000a}. Studies of one-dimensional Fr\"ohlich polaron models found excitations of the electron as resonances at high energies within the phonon continuum \cite{Devreese1964,Evrard1965}. In contrast, for the Holstein model of a small polaron, the existence of a bound excited state below the continuum is well known \cite{1974JPSJ...36..770S,1999PhRvB..60.1633B,2006PhRvB..74x5104G}.

In this work we calculate the energies of the bound excited states of the phonon dressing cloud for the Fr\"ohlich model in one dimension in the strong coupling limit, and provide numerical evidence for them in the weak to intermediate coupling regime applying Full Configuration Interaction Quantum Monte Carlo (FCIQMC) \cite{Booth2009a,Petruzielo2012,Blunt2015a}.
The first bound excited state appears at $\alpha\approx1.73$, which is in a crossover regime between weak and intermediate coupling strength with a spectral weight of $\approx 0.2$, suggesting that these states may be accessible in experiments.


FCIQMC is a modern flavor of Projector Quantum Monte Carlo \cite{Kalos2007}.
Originally developed for electronic structure calculations of molecules \cite{Booth2009a}, it has been extended to bosonic systems \cite{Yang2020,Brand2022a} and electron-phonon systems with infinite Hilbert space \cite{Anderson2022a}. Sampling the ground and excited \cite{Blunt2015a,Greene2022} states of the Fr\"ohlich model in Fock space provides us with direct access to the energies and properties of these states, avoiding the ill-conditioned analytic continuation procedure required to obtain spectral information from diagrammatic Monte Carlo calculations \cite{Mischenko2000a}.  Our state-of-the art semi\-stochastic \cite{Petruzielo2012,Blunt2015} implementation of FCIQMC \cite{RimuCodebase,RimuQMC} allows us to sample ground and excited states with moderate stochastic error and well-controlled systematic biases. Specifically, we report the first successful amelioration of the sign problem in Projector Monte Carlo \cite{Spencer2012a} by walker annihilation in an infinite Hilbert space without employing sign-problem-mitigating approximations.
Previously, this was only achieved in calculations with a finite Hilbert space \cite{Booth2009a,Shepherd2014,Kunitsa2020,Liebermann2022}.

\section{The model}
The Fröhlich model \cite{Fröhlich1954a} describes an electron in a polarizable medium.
Medium polarization is predominantly provided by longitudinal optical phonons \cite{Perebeinos2005,Devreese2009a}, and the one-dimensional model applies to the highly anisotropic situation of co-linear motion of electron and phonon motion \cite{Degani1986a}.
While the Coulomb interactions contributing to the coupling constant diverge for strictly one-dimensional systems \cite{Peeters1986a,Peeters1991a}, their contributions are regularized for quasi-one-dimensional quantum wires and nanotubes \cite{Haug1994,Gartstein2006,Gartstein2007}.
We parameterize the one-dimensional model with the dimensionless coupling constant $\alpha>0$ in the form \cite{Degani1986a,Peeters1991a}
\begin{align} \label{eq:Hamiltonian}
\hat{H} &= \frac{\hat{p}^2}{2m} + \hbar\omega_\mathrm{LO}\sum_k \hat{a}^\dagger_k \hat{a}_k \nonumber\\
&- \hbar\omega_\mathrm{LO}\sqrt{\frac{2\alpha}{L/l_0}} \sum_k \left(\hat{a}_k e^{ik\hat{r}} + \hat{a}^\dagger_k e^{-ik\hat{r}} \right),
\end{align}
where $\hat{p}$ and $\hat{r}$ are the electron momentum and position, $\hat{a}^\dagger_k$ and $\hat{a}_k$ are the phonon creation and annihilation operators, $\omega_\mathrm{LO}$ is the frequency of longitudinal optical phonons (taken to be constant), and $m$ is the band mass of the electron. $L$ is the length of the system with periodic boundaries (eventually taken to be infinite), and $l_0 = \sqrt{{\hbar}/{2m\omega_\mathrm{LO}}}$ is a natural unit of length. 

The energy-momentum spectrum consists of an isolated ground state band \cite{MR939707} with decreasing energy as the coupling strength increases, and a continuum of scattering states starting at an energy of $1\hbar\omega_\mathrm{LO}$ above the ground state \cite{Moller2006}. Although the absence of discrete excited states can be proven at  weak coupling~\cite{Seiringer2023}, 
discrete (bound) excited states with energies below the continuum exist at intermediate and strong coupling, as will be shown in the following.

\section{Excitation energies in the strong coupling limit}
In the limit of large $\alpha$, the discrete excited eigenvalues of $\hat{H}$ in Eq.~\eqref{eq:Hamiltonian} at total momentum $P=0$ can be obtained analytically,
and are given by
\begin{align} \label{eq:energy:expansion} 
   \frac{E_j}{\hbar\omega_\mathrm{LO}}\approx -\frac{\alpha^2} 3 - \frac{1}{2} \sum_{i\geq 0} (1 - \omega_i) + \omega_j, \quad j =0,1,\dots 
\end{align}
Since $\omega_j<1$ for all $j$, we thus expect that the number of discrete eigenvalues below the continuum increases without bounds as $\alpha\to \infty$, and this can actually be proved rigorously \cite{Mitrouskas2023a}.

As is well-known \cite{ADAMOWSKI1980249,MR709647,MR1462224}, the first term in Eq.~\eqref{eq:energy:expansion}, $-\alpha^2/3$, represents the semiclassical energy
obtained by minimizing the corresponding Pekar functional
with respect to the electron wave function $\psi(x)$ and the classical polarization field $\varphi(x)$. The Pekar functional (given in dimensionless units integrating over $x=r/l_0$) 
\begin{align}
 \mathcal{E}^{\rm P}[\psi , \varphi] = \int |\psi'|^2 - 2 \sqrt{2\alpha} \int |\psi|^2 \operatorname{Re} (\varphi) + \int |\varphi|^2 
\end{align}
is  obtained as an expectation value of the Hamiltonian of Eq.~\eqref{eq:Hamiltonian} with respect to a product of the electron wave function $\psi(x)$ and a coherent state for the phonon field with  amplitude $\varphi(x)$. The physical picture is that, at large coupling, the polarization field effectively becomes classical, and the electron occupies the ground state in the corresponding effective potential.
The Pekar functional has the explicit minimizer $\psi^{\rm P}(x) = \sqrt{\alpha/2} \,\mathrm{sech}(\alpha x)$ for the electron wave function and $\varphi^{\rm P}(x) = \sqrt{2\alpha} \psi^{\rm P}(x)^2$ for the polarization field.
The other terms in Eq.~\eqref{eq:energy:expansion} involving the ($\alpha$ independent) frequencies $\omega_j$ quantify the quantum fluctuations of the field around its classical value \cite{Tjablikow1954,Allcock01101956,allcock1963strong,MR4201293,MR4655801,Mitrouskas2023a}. Letting the electron take the instantaneous ground state in a varying field, they are obtained by expanding $\mathcal F^{\rm P} (\varphi ) = \min_\psi \mathcal{E}^{\rm P}[\psi, \varphi^{\rm P} + {\varphi}]$ to second order in ${\varphi}(x)$, with the result
\begin{align}
\label{classical:expansion}
& \mathcal F^{\rm P} (\varphi ) 
 \approx 
-\frac{\alpha^2}3 +  \| \text{Im} ({\varphi}) \|^2 + \langle \text{Re}({\varphi}),\hat \omega^2 \text{Re}({\varphi}) \rangle .
\end{align}
$\mathcal F^{\rm P} (\varphi )$ is equal to the sum of $\int |\varphi|^2$ and the ground state energy of the effective Hamiltonian $-\partial^2 -2\sqrt{2\alpha}\, \mathrm{Re}(\varphi)$, where $\partial = d/dx$. Accordingly, simple second order perturbation theory allows us to express the Hessian operator as
\begin{equation}
    \hat \omega^2 = 1 - 8 \alpha \psi^{\rm P} \frac {Q}{-\partial^2 - 2 \sqrt{2\alpha}\varphi^{\rm P} + \alpha^2}\psi^{\rm P},
\end{equation}
where  $\psi^{\rm P}(x)$ is understood as a multiplication operator, and  $Q$ projects orthogonal to $\psi^{\rm P}$. Note that $\psi^{\rm P}$ is the zero-energy ground state of $-\partial^2 - 2 \sqrt{2\alpha}\varphi^{\rm P} + \alpha^2 = A^\dagger A$ with $A=\partial - (\psi^{\rm P})'/\psi^{\rm P}$. We can write $(A^\dagger A)^{-1} Q = A^\dagger (A A^\dagger)^{-2} A$.  Since $A A^\dagger  = -\partial^2 + \alpha^2$, and $A \psi^{\rm P} = \psi^{\rm P} \partial$, the Hessian operator further simplifies to
\begin{align}\label{eq:hessian}
\hat{\omega}^2 =  1 + 8 \alpha \partial \psi^{\rm P}  ( -\partial^2 +  \alpha^2 ) ^{-2} \psi^{\rm P} \partial .
\end{align}

By scaling, the eigenvalues of $\hat\omega$ are independent of $\alpha$. 
We denote them  by  $0=\omega_0 < \omega_1 \leq \dots < 1$. There is a single zero mode, $\omega_0 = 0$, resulting from translation invariance. 
The eigenvalue equation $\hat\omega^2 f = \omega^2 f$ 
can be rewritten in terms of $\lambda \equiv 1- \omega^2$ 
as 
\begin{equation}\label{eq:fp}
    \left(\partial^2 - 4 \frac{(\psi^{\rm P})'}{\psi^{\rm P}} \partial+4\alpha^2 + \frac {8\alpha}\lambda(\psi^{\rm P})^2  \right) f' =0 .
\end{equation}
This follows from the identity $\psi^{\rm P} (-\partial^2  + \alpha^2)^2 (\psi^{\rm P})^{-1}= (\partial^2 - 4 \frac{(\psi^{\rm P})'}{\psi^{\rm P}} \partial + 4 \alpha^2 )\partial^2$. Changing variables from $x$ to $t= -\alpha^{-1} \psi^{\rm P}(x)' / \psi^{\rm P}(x) = \tanh(\alpha x)$ and making the ansatz $f'(x) = \ell(t)(1-t^2)$ turns Eq.~\eqref{eq:fp} into the standard Legendre equation
\begin{equation}\label{eq:Leg}
    (1-t^2) \ell'' - 2 t \ell' + n (n+1)\ell = 0 ,
\end{equation}
with $n(n+1) =  2 + 4\lambda^{-1}$. The general solution is thus a linear combination of $L^P_n$ and $L^Q_n$, the Legendre functions. The condition that the resulting function $f$ be square-integrable introduces the constraints $\int_{-1}^1 \ell(t) dt = 0$ and $\int_{-1}^1 \ell(t) \ln\frac{1+t}{1-t} dt = 0$ or, equivalently, $\int_{-1}^1 \ell(t) L_0^P(t) dt = 0 = \int_{-1}^1 \ell(t) L_0^Q(t) dt$ (due to the fact that $\int_{-\infty}^\infty f'(x) dx = 0$ and also $\int_{-\infty}^\infty f(x)dx = 0$ since $f$ itself is a derivative).

The even solutions of Eq.~\eqref{eq:Leg}, satisfying the orthogonality constraints, are simply the Legendre polynomials $L^P_n$ for $n=2,4,6,\dots$. This leads to (with $j \equiv n-2$)
\begin{equation}
    \omega_j^2 = 1 - \frac{4}{(j+4)(j+1)}\ , \quad j=0,2,4,\dots 
\end{equation}
In particular, $\omega_0 = 0$ and $\omega_2 = \sqrt{7/9}$. The odd solutions are slightly more complicated. They are linear combinations of $L^P_n$ and $L^Q_n$ for non-integer $n$. Using the well-known integrals over products of Legendre functions in \cite{MR698779}, the orthogonality conditions transform to the equation
\begin{equation}
    \psi_\Gamma(n+1) + \gamma = \frac \pi 2 \tan \frac{n \pi}{2}
\end{equation}
where $\psi_\Gamma = \Gamma'/\Gamma$ denotes the digamma function, and $\gamma \approx 0.577 $ is Euler's constant. There are infinitely many solutions, one in every interval $(2j, 2j+1)$ for $j\geq 1$, the smallest being $n_1\approx 2.523$. Denoting the solution in the interval $(2j, 2j+1)$ by $n_{2j-1}$, we thus conclude that
\begin{equation} \label{eq:oddexcited}
    \omega_j^2 = 1 - \frac{4}{(n_j+2)(n_j-1)}\ , \quad j=1,3,5,\dots 
\end{equation}
The lowest non-vanishing eigenvalues are given by $\omega_1\approx 0.647$ and $\omega_2=\sqrt{7/9} \approx 0.882$. The corresponding eigenstates alternate in parity, with the first excited state being even.

We proceed by quantizing the field $\varphi(x)\to \hat{\varphi}(x)$.
Expanding $\hat{\varphi}(x) = \sum_{j\geq 0} \hat{c}_j u_j(x)$ in terms of the (real-valued) eigenfunctions $u_j(x)$ of $\hat{\omega}^2$, and defining $\hat{p}_j = \sqrt{2} \text{Im}(\hat{c}_j)$ and $\hat{q}_j = \sqrt{2}\text{Re}(\hat{c}_j)$, we re-interpret the quadratic terms in Eq.~\eqref{classical:expansion} as a sum of harmonic oscillators with frequencies $\omega_j$:
\begin{align}
\label{eq:oscillator} 
\frac 12 \sum_{j \geq 0} \left( \hat{p}_j^2 + {\omega_j^2} \hat{q}_j^2\right). 
\end{align} 
From this expression we have to subtract the zero-point energy $1/2$ for every mode, since in the absence of the electron the field energy in Eq.~\eqref{eq:Hamiltonian} is normalized to have zero vacuum energy [i.e., $\sum_k \hat a^\dagger_k a_ k = \frac 12 \sum_{j\geq 0} ( \hat p_j^2 + \hat q_j^2 -1)$]. 
This leads to a contribution to the ground state energy of the form 
$\tfrac{1}{2} \sum_{i\geq 0} (\omega_{i} - 1)$, hence justifying Eq.~\eqref{eq:energy:expansion} for $j=0$. 
Numerically, one finds $\tfrac{1}{2} \sum_{i\geq 0} (\omega_{i} - 1) \approx - 0.955$. 
For $j \geq 1$, the excitation energies correspond to excitations of different oscillators in Eq.~\eqref{eq:oscillator}. Since $\omega_j > \tfrac{1}{2}$ for $j \geq 1$, only single excitations can appear below the continuum. The zero mode $j=0$ needs to be discarded for the calculation of the excitation spectrum of Eq.~\eqref{eq:Hamiltonian} due to the restriction to zero total momentum. 

 \section{Lee-Low-Pines transformation}
 Using a unitary transformation due to Lee, Low, and Pines amounting to a frame transformation \cite{LLP1953a}, the electron coordinate dependence of the Fr\"ohlich polaron Hamiltonian of Eq.~\eqref{eq:Hamiltonian} can be eliminated and the Hamiltonian rewritten in terms of the conserved total momentum $P$:
 \begin{align} \label{eq:LLP}
 \hat{H}_\mathrm{LLP} &= \frac{1}{2m}\left(P - \hbar \sum_k k \hat{a}^{\dagger}_k \hat{a}_k \right)^2 + \hbar\omega_\mathrm{LO} \sum_k \hat{a}^{\dagger}_k \hat{a}_k \nonumber\\
 & \quad - \hbar\omega_\mathrm{LO} \sqrt{\frac{2\alpha}{L/l_0}} \sum_k \left( \hat{a}^{\dagger}_k + \hat{a}_k \right).
 \end{align}
This equivalent Hamiltonian acts only on a Fock space of phonons for a given value of the total momentum $P$ and is more convenient for numerical calculations. 
We take the system size $L$ finite, which discretizes the phonon modes with the values of $k$ restricted to multiples of $2\pi/L$. A finite number of modes $M$ used in numerical calculations results in a momentum cutoff of $k_c = \pi(M-1)/L$. Using a larger number of modes increases the computational cost. Even so, the phonon Hilbert space dimension is still formally (and practically) infinite, as an arbitrary number of phonons can reside in each mode. Technically, the number of phonons in each mode is limited to be $<256$ by the 8-bit number type used for representing the Fock basis.
We perform calculations with up to $M=73$ and Hilbert space dimension $256^M\approx 6\times 10^{175}$.
In the following, we will restrict the analysis to $P=0$.

\section{FCIQMC} 
The algorithm samples the ground state of $\hat{H}_\mathrm{LLP}$ by repeatedly applying the operator $1 - \text{d}\tau(\hat{H}_\mathrm{LLP} - S)$ to a coefficient vector.
The shift parameter $S$ is updated 
to keep the norm 
(the number of `walkers') approximately constant \cite{Yang2020}, and therefore provides an estimator of the eigenstate energy after equilibration.
The operation is performed exactly for coefficients with a modulus above a given threshold and stochastically for the rest. 
Combined with stochastic vector compression \cite{Lim2017}, this implements a semistochastic algorithm, where part of the Hamiltonian matrix is treated deterministically while the rest is sampled stochastically \cite{Petruzielo2012,Blunt2015}. 
Crucially, contributions of opposite sign to a coefficient can annihilate and thus reduce sign noise. At sufficiently large walker number, the sign of the dominant coefficients becomes stable by a spontaneous symmetry-breaking process \cite{Spencer2012a,Shepherd2012b} and the sign problem is mitigated.
The residual noise leads to stochastic errors and 
also causes a small 
population control bias \cite{Umrigar1993,Ghanem2021,Brand2022a}, which we control 
by increasing the walker number until undetectable levels of the bias are seen that are smaller than the stochastic error bars.

If both the box size $L$ and the momentum cutoff $k_c$ are large enough, the results obtained using the discretized model approach the continuum limit.
We find that the energies converge exponentially in $L$, but only algebraically in $k_c$ (see Appendix~\ref{app:ground_state}).

\section{Ground state energy}
In the weak coupling limit, the ground state energy has been found analytically using perturbation theory \cite{Chen1994a,Peeters1991a} and the path integral formalism \cite{Degani1986a}.
For $P=0$, the ground state energy is
\begin{equation} \label{eq:weak}
\frac{E_0}{\hbar\omega_\mathrm{LO}} \approx -\alpha - 0.06066\alpha^2 - 0.00844\alpha^3 + \mathcal{O}(\alpha^4).
\end{equation}
While the numerical prefactors were rounded, this is exact to third order in $\alpha$.
In the strong coupling limit, the ground state energy of Eq.~\eqref{eq:energy:expansion} evaluates to
\begin{equation} \label{eq:strong}
\frac{E_0}{\hbar\omega_\mathrm{LO}} \approx -\frac{\alpha^2}{3} - 0.955 + \mathcal{O}(\alpha^{-2}).
\end{equation}

\begin{figure}
\includegraphics[width=\columnwidth]{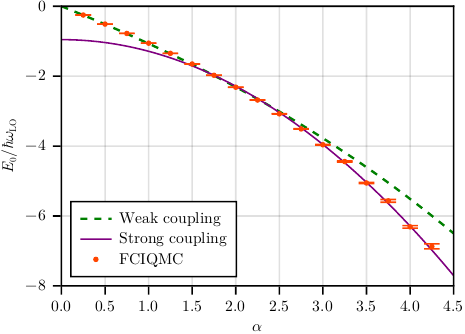}
\caption{The ground state energy $E_0$ of the one-dimensional Fröhlich polaron, calculated using FCIQMC for various
coupling strengths, compared with theoretical expansions in the weak [Eq.~\eqref{eq:weak}] and strong coupling [Eq.~\eqref{eq:strong}] regimes. The FCIQMC calculations used $L=6l_0$ and $k_c=12\pi/l_0$.
}
\label{fig:ground_state_energy}
\end{figure}

Figure \ref{fig:ground_state_energy} shows the transition from the weak coupling regime, through an intermediate coupling regime around $\alpha \approx 2$ where both expansions approximately overlap, into the strong coupling regime as the coupling strength $\alpha$ increases.
While there is no phase transition separating the weak and strong coupling regimes, different physical pictures apply \cite{Gerlach1991}. Our FCIQMC approach allows us to probe both regimes and the transition region comfortably. 
At $\alpha=2$ we estimate the discretization error to be around 1\% and about an order of magnitude larger than the stochastic error (see Appendix~\ref{app:ground_state}).

\section{Excited state energies} In order to calculate excited states numerically with FCIQMC, the computation is initialized with multiple starting vectors, which are then orthogonalized at every step (or a specified number of steps) of the computation using the Gram-Schmidt procedure.
This ensures that the $n^{\text{th}}$ vector converges to the $n^{\text{th}}$ eigenvector, and the $n^{\text{th}}$ shift acts as an estimator for the $n^{\text{th}}$-lowest eigenvalue~\cite{Blunt2015a}.

Even though the ground state calculation with the Hamiltonian \eqref{eq:LLP} is free of the Monte Carlo sign problem, excited state calculations introduce the sign problem \cite{Spencer2012a,Shepherd2014} due to the necessary existence of nodes in the excited state wave functions \cite{Blunt2015a}.
The sign problem manifests itself at low walker number in fluctuating signs of all state vector coefficients and an average value of the shift $\langle S \rangle$ that is lower than the correct energy \cite{Booth2009a,Liebermann2022}.
Increasing the walker number above a critical value, the shift average stabilizes and becomes independent of the walker number, indicating that the sign problem is overcome. For the studied range of  $\alpha$ values we find that $10^7$ walkers is sufficient, indicating that the sign problem is relatively weak in this case \cite{Liebermann2022}. This is remarkable as the unconstrained phonon occupation makes the Hilbert space of the Fr\"ohlich Hamiltonian \eqref{eq:LLP} infinite.
For details, see Appendix~\ref{app:sign_problem}.

\begin{figure}
\includegraphics[width=\columnwidth]{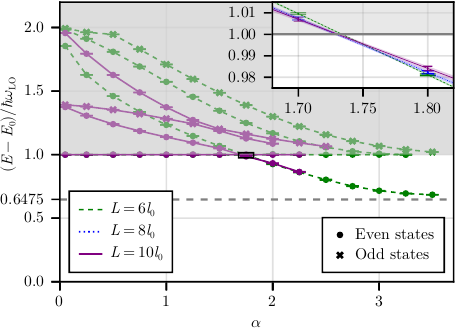}
\caption{
Energies of excited states relative to the ground state for the one-dimensional Fröhlich model with zero total momentum. The grey shaded region shows the phonon continuum starting $1\hbar\omega_\mathrm{LO}$ above the ground state. The dashed horizontal line indicates the strong-coupling prediction for the first bound excited state from Eq.~\eqref{eq:oddexcited}.
Symbols and error bars show finite system numerical results calculated using FCIQMC for two different box sizes ($L = 6l_0$, green, and $L = 10l_0$, purple) and classified by parity (circles for even and crosses for odd parity). The states inside the phonon continuum are discretized and strongly affected by the finite box size. A discretization-insensitive even state with energy below the phonon continuum for $\alpha \gtrapprox 1.73$ provides numerical evidence for a bound excited state of the phonon dressing cloud. The inset shows a detail near the emergence of the bound state and includes additional data for box size $L = 8l_0$. 
}
\label{fig:excited_states}
\end{figure}

Figure~\ref{fig:excited_states} shows the excitation energies of a few low-lying excited states as function of $\alpha$ and for different box sizes $L$ at  $P = 0$.
At the bottom of the phonon continuum (shaded region)
there is always one state with an excitation energy of $1\hbar\omega_\mathrm{LO}$. 
It is given by $|\mathrm{C}\rangle = ( \hat{a}_0^{\dagger} - \sqrt{{2\alpha l_0}/{L}})|\mathrm{GS}\rangle$, where $|\mathrm{GS}\rangle$ denotes the ground state. For weak coupling this is the 
first excited state in the spectrum.
Due to the finite box size, the continuum of eigenstates with excitation energies above $1\hbar\omega_\mathrm{LO}$ is discretized.
This discretizaton is most easily understood for vanishing coupling strength $\alpha = 0$. In this case, the Hamiltonian is diagonal in the Fock state basis with eigenvalues
${E_{a,b}}/{\hbar\omega_\mathrm{LO}} = a + (2\pi b l_0/L)^2$,
where $a \in \mathbb{N}$ is the total number of phonons, and $b \in \mathbb{Z}$ is the total phonon momentum in units of $\frac{2\pi\hbar}{L}$.
As the coupling strength $\alpha$ increases, 
some of the excitation energies decrease.

As seen in Fig.~\ref{fig:excited_states}, 
the first bound state with excitation energy below $1\hbar\omega_\mathrm{LO}$ appears at $\alpha \approx 1.73$, in the intermediate coupling regime. While the excitation energies in the discretized phonon continuum strongly depend on the box size $L$, this dependence disappears for the excited state below the continuum threshold.
The value of $\alpha$ where the first, even-parity bound state emerges was found by repeating calculations with different $L$ at $\alpha = 1.7$ and $\alpha = 1.8$, which resulted in the consistent interpolated crossing point shown in the inset of Fig.~\ref{fig:excited_states}. For $L = 10l_0$, the threshold was found to be $\alpha = 1.731 \pm 0.003$. The second bound state is of odd parity in the strong coupling limit, but not seen to breach the continuum threshold in the available numerical data. It is expected to appear at stronger coupling.

The excitation energies were found to converge more slowly in $L$ than the ground state energy, showing that the excited states are larger in real space than the ground state.
However, the excitation energies converge more quickly in the momentum cutoff $k_c$ than the absolute energies, and $k_c = 4\pi/l_0$ is large enough to find accurate excitation energies in the range of $\alpha$ reported here. For the bound excited state, the energy is converged at $L=6l_0$ and $k_c = 4\pi/l_0$.
See Appendix~\ref{app:excited_states} for more details.

\section{Spectral weight} 
\begin{figure}[h!]
\includegraphics[width = \columnwidth]{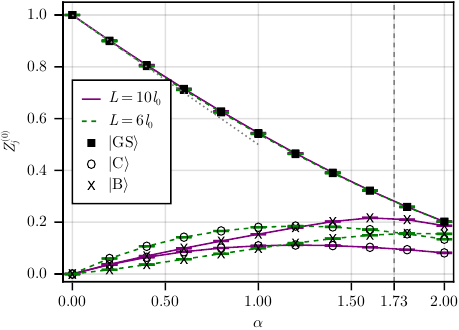}
\caption{Spectral weight $Z_j^{(0)}$ of the three lowest energy states vs.~the coupling strength $\alpha$ for $k_c=4\pi/l_0$ and two different box sizes. For $\alpha> 1.73$ marked by the vertical dashed line
the state $|\mathrm{B}\rangle$ becomes a bound excited state.
The dotted line is the first order perturbation theory result $1-\alpha/2$ for the ground state.
}
\label{fig:spectral}
\end{figure}

The spectral weight of the state $|j\rangle$ at the total momentum $P$ from the Hamiltonian \eqref{eq:LLP}
\begin{align}
   Z_j^{(P)} = |\langle \mathrm{vac}|j \rangle|^2 ,
\end{align}
carries important information about the ability to detect the state in spectroscopic measurements (e.g.~angle-resolved photoemission spectroscopy (ARPES) \cite{Berglund1964,Koyama1970,Mishchenko2018,Nery2018} or polaron injection spectroscopy \cite{Jorgensen2016,Hu2016,Vale2021a}). Figure~\ref{fig:spectral} shows the spectral weight of the three lowest states in the discretized model. While the spectral weight of the ground state decreases $\sim 1-\alpha/2$ for small $\alpha$ as expected \cite{Mischenko2000a}, the bound excited state carries a significant spectral weight $\approx 0.2$ near the threshold. The spectral weight of states within the continuum ($|\mathrm{C}\rangle$, and $|\mathrm{B}\rangle$ for $\alpha<1.73$) changes with box size and is expected to vanish in the limit $L \to \infty$. The spectral weight of the discrete states ($|\mathrm{GS}\rangle$, and $|\mathrm{B}\rangle$ for $\alpha>1.73$) is expected to depend weakly on the discretization of the model but a full convergence study is yet to be performed.

The number of phonons of the discrete states, $N_{|\mathrm{GS}\rangle} \approx 1.94$ and $N_{|\mathrm{B}\rangle} \approx 3.68$ at $\alpha=2$ (and $L=6l_0$, $k_c=4\pi/l_0$), reveals that extra phonons are bound by the polaron in the first excited state. This number is larger than in the strong coupling limit $\alpha\to\infty$ where $N_{|\mathrm{B}\rangle} - N_{|\mathrm{GS}\rangle} = 0.6475$. Plots of the phonon densities are shown in Appendix~\ref{app:phonon_density}.

\section{Outlook}
An outstanding question concerns the relevance of our findings to materials \cite{Franchini2021} or experimentally accessible synthetic quantum systems \cite{Massignan2025,Grusdt2025}.
Polaron effects are expected to be relevant in various quasi-one-dimensional settings, such as carbon nanotubes \cite{Perebeinos2005,Verissimo-Alves2001,Gartstein2006,Gartstein2007}, quantum wires \cite{Maslov2010}, or polymer chains \cite{Franchini2021}.
Being a foundational and highly simplified model, however, the Fröhlich polaron model typically has to be extended to be applied to realistic materials \cite{Franchini2021}.
Recently predicted discrete excited phonon branches near a Mott-insulator--superfluid phase transition in a Bose-Hubbard model \cite{Alhyder2025} may be a related phenomenon to the excited states studied in this work. An interesting avenue of future research will be to extend the FCIQMC approach to the three-dimensional Fr\"ohlich model to find the threshold coupling constant for the bound excited state and independently check the diagrammatic Monte Carlo results of Ref.~\cite{Mischenko2000a}, which indicated the absence of bound excited states up to values of $\alpha \approx 8$. The question remains as to the threshold for the appearance of the second, and further, bound excited states. Due to the large computational cost of large-$\alpha$ simulations, approaches such as importance sampling may be needed to improve the efficiency of the method in order to find these thresholds. It will further be valuable to understand the class of problems in which the resource requirements for sign-problem amelioration by walker annihilation in projector quantum Monte Carlo decouple from the dimension of Hilbert space.

\begin{acknowledgments}
We are grateful to Dmytro Kolisnyk for his help in working out the spectrum of the Hessian. 
This work was supported by the Marsden Fund of New
Zealand (contract no.\ MAU2007) from government funding
administered by the Royal Society Te Ap\=arangi and by a summer scholarship from Te Whai Ao -- Dodd-Walls Centre for Photonic and Quantum Technologies and the Physics Department, University of Auckland.
We acknowledge support by the New Zealand eScience Infrastructure (NeSI) high-performance computing facilities in the form of a merit project allocation.

\section*{Data availability}
All numerical calculations were performed using the open source software \texttt{Rimu.jl} \cite{RimuQMC}. The data that support the findings of this article are openly available \cite{zenododata}.
\end{acknowledgments}

\appendix

\section{\label{app:ground_state}Ground State Energy Convergence}
The energy of the ground state calculated using our computational model
has a systematic discretization error, with the energy becoming more accurate as
the box size $L$ and the momentum cutoff $k_c$ are increased.

The results shown here use a coupling strength of $\alpha = 2$. For different $\alpha$
the overall behavior is the same, with the errors caused by finite $L$ and $k_c$ being roughly proportional to the 
ground state energy.

\begin{figure}[!htb]
\includegraphics[width=\columnwidth]{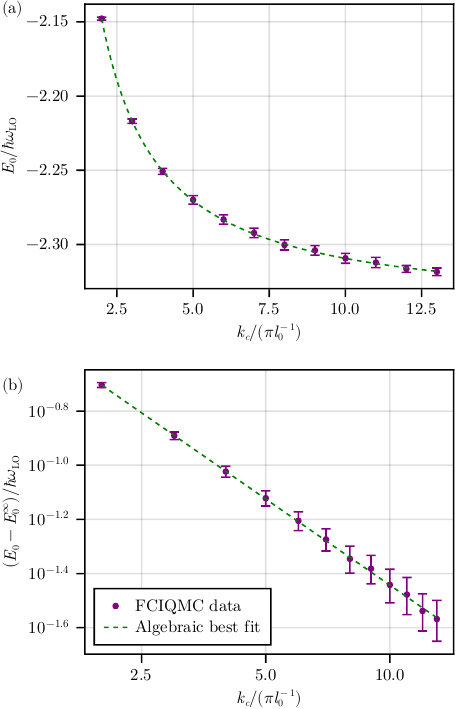}
\caption{(a) The ground state energy for $\alpha = 2$ with $L = 6l_0$ as a function of the cutoff $k_c$. (b) The difference between the ground state energy and the extrapolated energy at infinite $k_c$, on a log-log plot. The trend suggests an algebraic convergence of the energy, with an error in the energy proportional to approximately $k_c^{-1}$.
}
\label{fig:ground_state_kc}
\end{figure}

Keeping $L$ constant, the ground state energy decreases with increasing momentum cutoff $k_c$ consistent with algebraic convergence, as shown in Fig.~\ref{fig:ground_state_kc}.
The best fit curve is 
\begin{align}
E_0&/(\hbar\omega_\mathrm{LO}) = -2.3455 \pm 0.0039 \nonumber\\
&+ \left[( 0.4110 \pm 0.0070  ) \frac{k_c l_0}{\pi} \right]^{(-1.0568 \pm 0.0451)}.
\end{align}
The exponent was found to be close to $-1$ for other values of $\alpha$ as well.

Using this, we can estimate that the error incurred in the ground state energy from limiting the maximum phonon wavenumber to $k_c=12\pi/l_0$ is $(0.0297 \pm 0.0064)\hbar\omega_\mathrm{LO}$, about $1.3\%$ of the energy.
For $k_c = 4\pi/l_0$ the error is $(0.095 \pm 0.0082)\hbar\omega_\mathrm{LO}$, about $4.1\%$ of the energy.

\begin{figure}[!htb]
\includegraphics[width=\columnwidth]{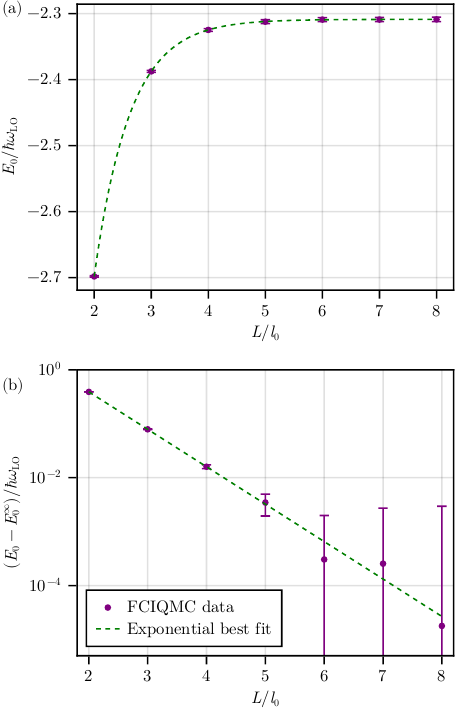}
\caption{(a) The ground state energy for $\alpha = 2$ with $k_c=10\pi/l_0$, for various box sizes $L$. (b) The difference between the ground state energy and the extrapolated energy at infinite $L$, on a semi-log plot. In this case, the data imply an exponential convergence of the energy.}
\label{fig:ground_state_l}
\end{figure}

Figure~\ref{fig:ground_state_l} suggests an exponential convergence of the energy in the box size $L$ when $k_c$ is kept constant.
The best fit curve is 
\begin{align}\label{eqn:ground_state_l_convergence}
E_0&/(\hbar\omega_\mathrm{LO}) = -2.3089 \pm 0.0008 \nonumber\\
&- (9.5077 \pm 0.2381)\exp\left[-(1.5976 \pm 0.0132)\frac{L}{l_0}\right].
\end{align}
Note that the value at infinite $L$ is too high since the momentum cutoff is finite. The error caused by the finite box size at $L = 6l_0$ can be estimated from the observed exponential convergence to be $(0.00065 \pm 0.0011)\hbar\omega_\mathrm{LO}$, consistent with no error.

The calculations reported in Fig.~\ref{fig:ground_state_energy} were performed at $L=6l_0$ and $k_c = 12\pi/l_0$. 
These calculations with $M=73$ modes used $2 \times 10^5$ walkers to probe a Hilbert space with dimension $256^M\approx 6\times 10^{175}$. The number of non-zero vector coefficients was strongly dependent on $\alpha$. For $\alpha=0.25$ there were about $1.4 \times 10^4$ non-zero coefficients, whereas for $\alpha = 2$ there were about $1.8 \times 10^5$, almost as many as the number of walkers. Evolving for about $10^6$ steps resulted in a stochastic error in the energy estimate of approximately $3 \times 10^{-3}\hbar\omega_\mathrm{LO}$ for $\alpha = 2$. This is an order of magnitude smaller than the estimated discretization error due to the finite value of $k_c$.

\section{\label{app:excited_states}Excited State Energy Convergence}
\subsection{Bound Excited State}
\begin{figure}[htb]
\includegraphics[width=\columnwidth]{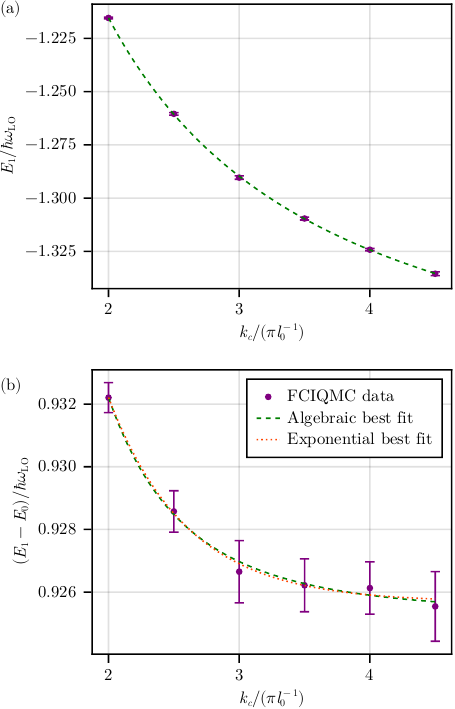}
\caption{(a) Energy of the first excited state for $\alpha = 2$ and $L = 6l_0$, with varying momentum cutoff. There is an apparent algebraic convergence, similar to that of the ground state energy. (b) The energy difference between the bound excited state and the ground state for $\alpha=2$ and $L=6l_0$, with varying $k_c$. The energy difference converges more quickly than the energies of the states separately.}
\label{fig:excited_state_kc}
\end{figure}

At $\alpha = 2$, the first excited state is discrete corresponding to an excitation of the phonon cloud that is bound to the electron (polaron). The energy of this state converges in $k_c$ in a similar way to the ground state indicating algebraic convergence. See Fig.~\ref{fig:excited_state_kc}(a). The best fit curve (at $L = 6l_0$) is
\begin{align}
E_1&/(\hbar\omega_\mathrm{LO}) = -1.4082 \pm 0.0046 \nonumber\\
&+\left[ (0.4429 \pm 0.0046)\frac{k_c l_0}{\pi} \right]^{-1.2002 \pm 0.0576}.
\end{align}

Although the excited state energy has a similarly slow convergence with $k_c$ as the ground state, we find that the difference between the two, i.e.~the excitation energy, converges much more quickly in $k_c$, see Fig.~\ref{fig:excited_state_kc}(b). 
Therefore, even though there is a large error in the energies of the ground state and first excited state separately
with a momentum cutoff of $k_c = 4\pi/l_0$, this $k_c$ is large enough to find the excitation energy.
With a momentum cutoff this small, the energies of the ground state and the excited states are too high by about $0.1\hbar\omega_\mathrm{LO}$, but the differences between the states are constrained mostly to the low momentum modes resulting in more precise excitation energies. 
While the numerical data is not accurate enough to distinguish between algebraic and exponential convergence, both extrapolations estimate a systematic error at $k_c = 4\pi/l_0$ that is smaller than the stochastic error bar and consistent with no error.    

\begin{figure}[!htb]
\includegraphics[width=\columnwidth]{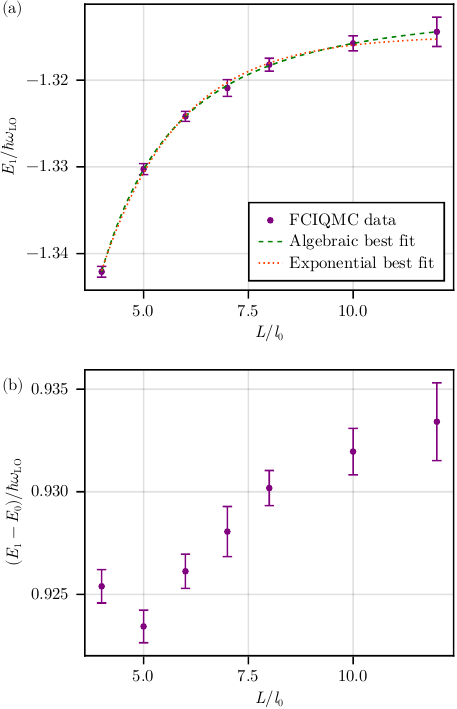}
\caption{(a) Energy of the first excited state for $\alpha = 2$ and $k_c = 4\pi/l_0$, with varying box size. The rate of convergence is unclear, but even with a slow algebraic convergence the box size error is on the order of $0.01\hbar\omega_\mathrm{LO}$. (b) The excitation energy of the bound state with the same parameters. This is not monotonic because the energies of the two states converge differently.
}
\label{fig:excited_state_l}
\end{figure}

On the other hand, Fig.~\ref{fig:excited_state_l} shows that the energy of the first excited state converges more slowly in $L$ than the ground state, which was already
converged at $L = 6l_0$. With an exponential fit, the exponent is $(-0.54 \pm 0.05)L/l_0$, indicating a slower convergence than the ground state \eqref{eqn:ground_state_l_convergence}.

Assuming instead an algebraic convergence for the worst case scenario, the error in the excitation energy caused by the finite box size at $L = 6l_0$ estimated from extrapolation is
$(0.0100 \pm 0.0018)\hbar\omega_\mathrm{LO}$.

\subsection{Continuum Excited States}

The excited states we find in the continuum have excitation energies
larger than $1\hbar\omega_\mathrm{LO}$ with a $1/L$ energy dependence only because our computational model
is discretized. In the limit $L \to \infty$, all excitation energies of the individual discrete states shown in Fig.~\ref{fig:excited_states} would converge to $1\hbar\omega_\mathrm{LO}$. Note that this means that the linear
interpolation used to find the crossover point for the bound excited
state (see Fig.~\ref{fig:excited_states} inset) is only valid up to a certain precision. To find the
crossing point more precisely with large $L$, one would need
to take more data at smaller intervals of $\alpha$.

\section{\label{app:sign_problem}Overcoming the sign problem in excited state FCIQMC}

\begin{figure}
\includegraphics[width=\columnwidth]{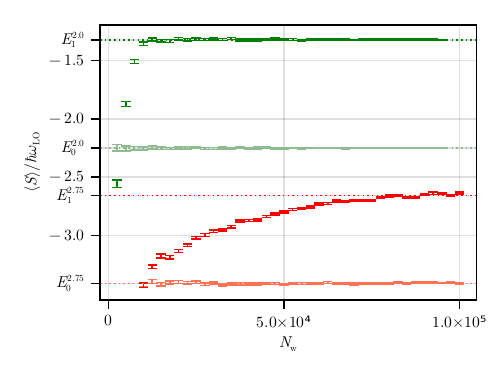}
\caption{Overcoming the sign problem by walker annihilation. The mean of the shift energy $\langle S\rangle$ is shown as a function of the number of walkers $N_\mathrm{w}$ for the ground and first excited states at two values of the interaction strength $\alpha = 2.0$ (green) and $\alpha = 2.75$ (red), with $L=6l_0$ and $k_c=4\pi/l_0$. The critical number of walkers $N_c$ above which the mean shift becomes an estimator for the excited state energy depends on the interaction strength $\alpha$.}
\label{fig:sign_problem}
\end{figure}

The Hamiltonian of the Fr\"ohlich polaron of Eq.~\eqref{eq:LLP} is stoquastic, i.e.~all non-zero off-diagonal elements are negative.
Thus, it avoids the sign problem for the ground state, as stochastic spawns never change the sign of a walker \footnote{Specifically, all matrix elements of the operator $1 - \text{d}\tau(\hat{H}_\mathrm{LLP} - S)$ that defines the FCIQMC algorithm are non-negative for small enough $\text{d}\tau$, and a ground state vector with a uniform sign is obtained upon iteration.}. This changes for the calculation of excited states where the Gram-Schmidt projector introduces sign-changing contributions to coefficients. In the absence of walker annihilation or at a small number of walkers where annihilation is inefficient, the iterations converge to a ground state of a different Hamiltonian with a lower (incorrect) ground state energy \cite{Spencer2012a}. This leads to an incorrect value of the mean shift energy $\langle S \rangle$. 
Typically, the presence of a sign problem in Projector Monte Carlo requires a suppression mechanism that introduces uncontrolled approximations (e.g.\ fixed nodes \cite{Reynolds1982a}, a constrained path \cite{Zhang1997}, the phaseless \cite{Zhang2003} or initiator method \cite{Cleland2010}). 
Formulated in a discrete Hilbert space, FCIQMC can overcome the sign problem by walker annihilation, avoiding such approximations, if the coefficient vector is sufficiently strongly localized \cite{Booth2009a,Spencer2012a,Shepherd2014}.

Figure \ref{fig:sign_problem} shows how the mean shift energy depends on the number of walkers for the ground and the first excited state of the Fr\"ohlich polaron problem.
For the ground state, 
the mean shift only carries small stochastic fluctuations and is a valid estimator for the ground state energy regardless of the walker number because the sign problem is absent. The population control bias, which raises $\langle S \rangle$ above the correct value for small walker numbers \cite{Brand2022a}, is smaller than the stochastic errors in this case. For the excited states, the strong dependence of the shift energy on the walker number below a critical value indicates the presence of a sign problem. Above the critical walker number $N_\mathrm{c}$, the mean of the shift fluctuates around the correct value of the excited state energy.
This critical number is found to be strongly dependent on the value of the coupling strength $\alpha$. We find values of around $N_\mathrm{c} \approx 1\times 10^4$ for $\alpha=2.0$ and $N_\mathrm{c} \approx 1\times 10^5$ for $\alpha=2.75$,  as seen from Fig.~\ref{fig:sign_problem}. The calculations reported in Fig.~\ref{fig:excited_states} and Fig.~\ref{fig:spectral} were obtained with $N_\mathrm{w}$ on the order of $10^6$ or $10^7$, at least an order of magnitude higher than the observed critical values $N_\mathrm{c}$ for overcoming the sign problem.

\section{\label{app:phonon_density}Phonon Density}
\begin{figure}
\includegraphics[width=\columnwidth]{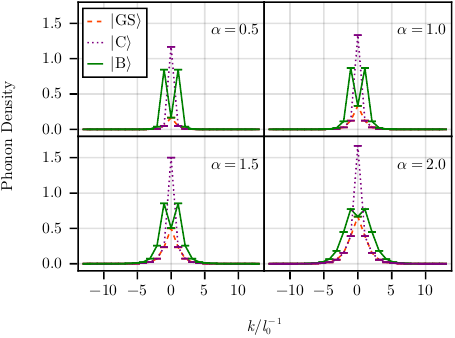}
\caption{The density of virtual phonons in each mode  $\langle j| a_k^\dagger a_k |j \rangle$ for the ground state $|j\rangle =|\mathrm{GS}\rangle$ and first two excited states, with $L=6l_0$ and $k_c = 4\pi/l_0$, for various coupling strengths. The state $|\mathrm{C}\rangle$ is at the start of the continuum, and always has an excitation energy of $1\hbar\omega_\mathrm{LO}$. The state $|\mathrm{B}\rangle$ is a bound excited state for $\alpha \gtrsim 1.73$ and otherwise part of the discretized continuum. For the total phonon numbers see Table \ref{tab:phonons}.}
\label{fig:densities}
\end{figure}

The bound excited state can be studied more closely by looking at the density of virtual phonons in each mode, as shown in Fig.~\ref{fig:densities}.
The ground state gains more phonons as the coupling strength increases, with a larger density in the lower momentum modes. The continuum threshold state $|\mathrm{C}\rangle$ has increased phonon density only in the $k=0$ mode, as expected.

For weak coupling, the second excited state is part of the discretized continuum and mostly a combination of the Fock state with one phonon in each of the $k = \pm 2\pi/L$ modes (one of the $a = 2$, $b = 0$ states), and the two Fock states with one phonon, in each of those modes respectively (the $a = 1$, $b = \pm 1$ states). So for weak coupling the state can be approximated as $\left[ c_1\left(\hat{a}^{\dagger}_{2\pi/L} + \hat{a}^{\dagger}_{-2\pi/L}\right) + c_2\hat{a}^{\dagger}_{2\pi/L}\hat{a}^{\dagger}_{-2\pi/L} \right]|\text{vac}\rangle$ where $|\text{vac}\rangle$ is the phonon vacuum, for some $c_1$ and $c_2$ depending on $L$. As the coupling strength increases, there is a larger occupation in the other modes. For $\alpha = 2$, when this state drops below the continuum to become a bound excited state, 
this simple description no longer applies. The bound state phonon density then becomes much more weakly affected by the mode discretization and the box size.

The number of phonons in each of the states in Fig.~\ref{fig:densities} is presented in Table \ref{tab:phonons}.

\begin{table}[ht]
    \centering
    \setlength{\tabcolsep}{10pt}
    \renewcommand{\arraystretch}{1.4}
   \begin{tabular}{c | c c c}
    $\alpha$ & $N_{|\mathrm{GS}\rangle}$ & $N_{|\mathrm{C}\rangle}$ & $N_{|\mathrm{B}\rangle}$\\
    \hline
    0.5 & 0.29 & 1.29 & 1.95\\
    1.0 & 0.67 & 1.67 & 2.36\\
    1.5 & 1.19 & 2.19 & 2.96\\
    2.0 & 1.94 & 2.94 & 3.68
    \end{tabular}
    \caption{The total number of phonons in the dressing cloud as the integral of the phonon density for the states presented in Fig.~\ref{fig:densities}. The continuum threshold state $|\mathrm{C}\rangle$ contains one more phonon than the ground state. In the strong coupling limit $\alpha\to\infty$, the first bound excited state $|\mathrm{B}\rangle$ has 0.6475 more phonons than the ground state.}
    \label{tab:phonons}
\end{table}

\bibliography{polaron}

\end{document}